\documentclass[aps,prd,twocolumn,superscriptaddress,showpacs,showkeys]{revtex4-1}
\usepackage{graphicx,color}
\usepackage{rotating}
\usepackage{amssymb}
\usepackage{amsfonts}
\usepackage{amsmath}

\usepackage{listings}
\makeindex

\usepackage[super]{nth}
\usepackage{braket}

\newcommand{\be}{\begin{equation}}
\newcommand{\ee}{\end{equation}}
\newcommand{\bea}{\begin{eqnarray}}
\newcommand{\eea}{\end{eqnarray}}

\newcommand\D{\mathrm{d}}

\begin{document}


\title{Extended Uncertainty Principle for Rindler and cosmological horizons\\ 
} 




\author{Mariusz P. D\c{a}browski}
\email[]{mariusz.dabrowski@usz.edu.pl}
\affiliation{Institute of Physics, University of Szczecin, Wielkopolska 15, 70-451 Szczecin, Poland}
\affiliation{National Centre for Nuclear Research, Andrzeja So{\l}tana 7, 05-400 Otwock, Poland}
\affiliation{Copernicus Center for Interdisciplinary Studies, Szczepa\'nska 1/5, 31-011 Krak\'ow, Poland}

\author{Fabian Wagner}
\email[]{fabian.wagner@usz.edu.pl}
\affiliation{Institute of Physics, University of Szczecin, Wielkopolska 15, 70-451 Szczecin, Poland}
\date{\today}
\begin{abstract}
We find exact formulas for the Extended Uncertainty Principle (EUP) for the Rindler and Friedmann horizons and show that they can be expanded to obtain asymptotic forms known from the previous literature. We calculate the corrections to Hawking temperature and Bekenstein entropy of a black hole in the universe due to Rindler and Friedmann horizons. The effect of the EUP is similar to the canonical corrections of thermal fluctuations and so it rises the entropy signalling further loss of information. 


\end{abstract}

\pacs{}
\keywords{}

\maketitle

\section{Introduction}

The Heisenberg Uncertainty Principle (HUP) constitutes a cornerstone of the quantum physics and is rooted in the quantisation of electromagnetic radiation resulting in photons. It introduces the Planck fundamental constant $\hbar$ which together with two other fundamental constants - the speed of light $c$, and Newton's gravitational constant $G$ - form the Planck (natural) scale in physics. One of the units of this scale is the Planck length $l_p = \sqrt{G \hbar/c^3}$. In fact, the HUP does neither take into account quantum gravity effects of the photon interaction nor the curvature of space-time. 
However, widely spread by the 
considerations of superstring theory  \cite{stringa,stringb,stringc,stringd,stringe} and loop quantum gravity \cite{lqg1,lqg2},  these two phenomena have been gradually taken into account resulting in the Generalised Uncertainty Principle (GUP)  \cite{GUP2,GUP3,GUP4,GUP5,Ghosh,Leandros,alpha2,Calmet,GUPReview,Sparsity,GUP-Unruh} and the Extended Uncertainty Principle (EUP) \cite{Mignemi2010,Anrade,Mureika,EUPThermod}, or were even put together as the Generalised Extended Uncertainty Principle (GEUP) \cite{AdlerDuality,Bambi2008,MIPark,Bolen2005,Zhu2009}. Yet, a different approach to the problem using the so-called qmetric was also considered \cite{Kothawala}.

In terms of the standard deviations of position and  momentum
\begin{align}
\sigma_x^2=\braket{\hat{x}^2}-\braket{\hat{x}}^2\\
\sigma_p^2=\braket{\hat{p}^2}-\braket{\hat{p}}^2
\end{align}
and in the context of space-times with external horizons, the most general asymptotic form of the GEUP which includes both the GUP and the EUP can be formulated as \cite{AdlerDuality,Bambi2008} 
\be
\sigma_x \sigma_p \geq \frac{\hbar}{2} \left(1 + \frac{\alpha_0 l_p^2}{\hbar^2} \sigma_p^2 +  \frac{\beta_0}{r_{hor}^2} \sigma_x^2\right),
\label{GEUP} 
\ee
where $x$ is the position, $p$ the momentum, $l_p$ plays the role of the minimum length, $r_{hor}$ is the radius of the horizon which is introduced by the background space-time, and $\alpha_0, \beta_0$ are  
dimensionless parameters. 

An interesting property of (\ref{GEUP}) which relates to superstring theory \cite{Green} is the invariance of it under the (duality) transformations
\be
 \frac{\sqrt{\alpha_0} l_p}{\hbar}  \sigma_p \leftrightarrow \frac{\hbar}{\sqrt{\alpha_0} l_p} \sigma_p^{-1} ,\hspace{0.3cm}   \frac{\sqrt{\beta_0}}{l_H}  \sigma_x \leftrightarrow \frac{l_H}{\sqrt{\beta_0}} \sigma_x^{-1} ,
\ee
just for the GUP sector ($\beta_0 =0$) and the EUP sector ($\alpha_0 = 0$) respectively, and 
\be
 \frac{\sqrt{\alpha_0} l_p}{\hbar}  \sigma_p \leftrightarrow \frac{\sqrt{\beta_0}}{l_H}  \sigma_x
\ee
for both sectors simultaneously. It is interesting to note some general relations between black hole and cosmological horizons \cite{Mielczarek}. 

There have been various derivations of the GUP which account for the gravitational part of the interaction between an electron and a photon including simple Newtonian arguments \cite{AdlerDuality}. The changes caused by $classical$ gravity could in principle have a great deal of implications. An example is the disappearance of the Chandrasekhar limit \cite{Rashidi} under the GUP and its recovery under the application of the EUP \cite{OngYao}. In fact, it emerges that the curvature effect is missing in the GUP and once the EUP is applied, 
it helps to recover the limit which is an observational fact. However, the most important consequence of the GUP is its influence on the Hawking temperature \cite{hr1a}  and Bekenstein entropy~\cite{bentropy}. In fact, it modifies the black hole evaporation process which ceases under GUP conditions leaving a remnant which stores information \cite{chen,rabin}  giving a possible solution to the information puzzle \cite{hr3}.

There have been some attempts to bound the GUP parameter $\alpha_0$ in (\ref{GEUP}) observationally\cite{alpha2011,alpha2018,alphaGW, alpha2018a,Bounds-beta} including the issue of its positivity or negativity \cite{alpha1,alpha2,alphaposit,alphaposit2,alphaposit3,alpha2011,alpha2018,alphaGW, alpha2018}. In fact, the microcanonical corrections reduce the entropy and so the parameter $\alpha_0$ seems to be negative while for the canonical corrections it should be positive \cite{AdlerDuality,alpha1,Sparsity}. It is also worth mentioning that there is some analogy between the GUP in particle physics and the solid state phenomena in graphene which could pave the 
path to experimentally support this idea \cite{Graphene}.  

A very nice, quite rigorous derivation of the EUP based on space-times of constant curvature was presented in Ref. \cite{Schuermann2018} and it directly shows that even classical gravity alters the uncertainty relation. This had been suggested earlier in the context of geometry and topology \cite{Golovnev,Schuermann}. In particular, if we make any measurement, we are certain that the particle we measure is located inside its own universe which thus restricts the uncertainty. 

The paper is organized as follows. 
In Section \ref{EUPgeom} we present the method of derivation of the EUP from geometrical arguments.  In Sections \ref{ERind} and \ref{EFried} we present the application of the method for Rindler space and Friedmann universes sliced in a way that the cosmological horizon appears manifestly. Section \ref{HawkingRadiation} describes a way to interpret some of the results as manifestations of Hawking radiation. In Section \ref{BHthermo} we discuss the influence of cosmological horizons onto the Hawking radiation and Bekenstein entropy of a local black hole.  


\section{Background geometry determined EUP}
\label{EUPgeom}

The underlying idea of our approach is that the measurement of momentum depends on a given space-time background \cite{Schuermann,Schuermann2018}. In order to measure the momentum one needs to consider a compact domain $D$ with boundary $\partial D$ characterised by the geodesic length $\Delta x$ around the location of the measurement with Dirichlet boundary conditions. Thus the wavefunction is confined to $D$. Note that $D$ lies on a spacelike hypersurface. Thus like other quantum gravity effects this method is observer dependent. The method then reduces to the solution of an eigenvalue problem for the wave function $\psi$:
\begin{align}
\hat{\Delta} \psi + \lambda \psi=0
\end{align} 
inside $D$ with the requirement that $\psi=0$ on the boundary $\partial D$, $\lambda$ denotes the eigenvalue, and $\hat{\Delta}$ is the Laplace-Beltrami operator. As we can choose $\psi$ to be real (the eigenvalue problem is the same for the real and the imaginary part), the Dirichlet boundary conditions assure that $\braket{\hat{p}}=0$, and so one can obtain the uncertainty of a momentum $\hat{p} = -i \hbar \partial_i$ measurement as
\begin{align}
\label{Dpgen}
\sigma_p=\sqrt{\braket{\hat{p}^2}} =\hbar\sqrt{-\braket{\psi|\hat{\Delta}|\psi}} \geq \hbar\sqrt{\lambda_1}
\end{align}
where $\lambda_1$ denotes the first eigenvalue. Multiplying by $\Delta x,$ the uncertainty relation corresponding to this momentum measurement is obtained. It was found for Riemannian 3-manifolds of constant curvature $K$ that \cite{Schuermann2018} 
\begin{align}
\label{Kgeom}
\sigma_p \Delta x\geq \pi\hbar \sqrt{1-\frac{K}{\pi^2}(\Delta x )^2} .
\end{align}
Note that the uncertainty relation derived this way is not of the same kind as the one described by \eqref{GEUP} because it features the characteristic length of confinement $\Delta x.$ The domain applied in this letter is a ball of radius $\Delta x.$ Thus, $\Delta x$ should rather be interpreted as uncertainty and does not describe the standard deviation of position.

\section{EUP for Rindler spacetime}
\label{ERind}

The method requires a foliation of spacetime into hypersurfaces of constant time and so we consider only the spatial part of the Rindler metric which is of the form 
\begin{align}
\label{Rmetric}
\D s^2=\frac{c^2\D l^2}{2 \alpha l}+\D \vec{y}_\perp^2
\end{align}
with the acceleration $\alpha$ describing a boost in the $l$-direction as applied to Minkowski space, $c$ the speed of light, and $\vec{y}_{\perp}$ denote all components of the metric perpendicular to $l-$direction. An observer (the measured particle) moving with the acceleration $\alpha$ is located at $l_0=2c^2/\alpha$ and sees a horizon at a distance $l_0$ at $l=0$. 

For simplicity the directions transversal to the acceleration will not play any role in this treatment. Thus, the obtained uncertainty will account for the effect on measurements done along the direction of acceleration.

As we are basically describing a one-dimensional problem, the domain can most conveniently be taken to be the interval $I=[l_0-\Delta x, l_0+\Delta x].$ 

\begin{figure}[!htb]
\centering
\includegraphics[width=.83\linewidth]{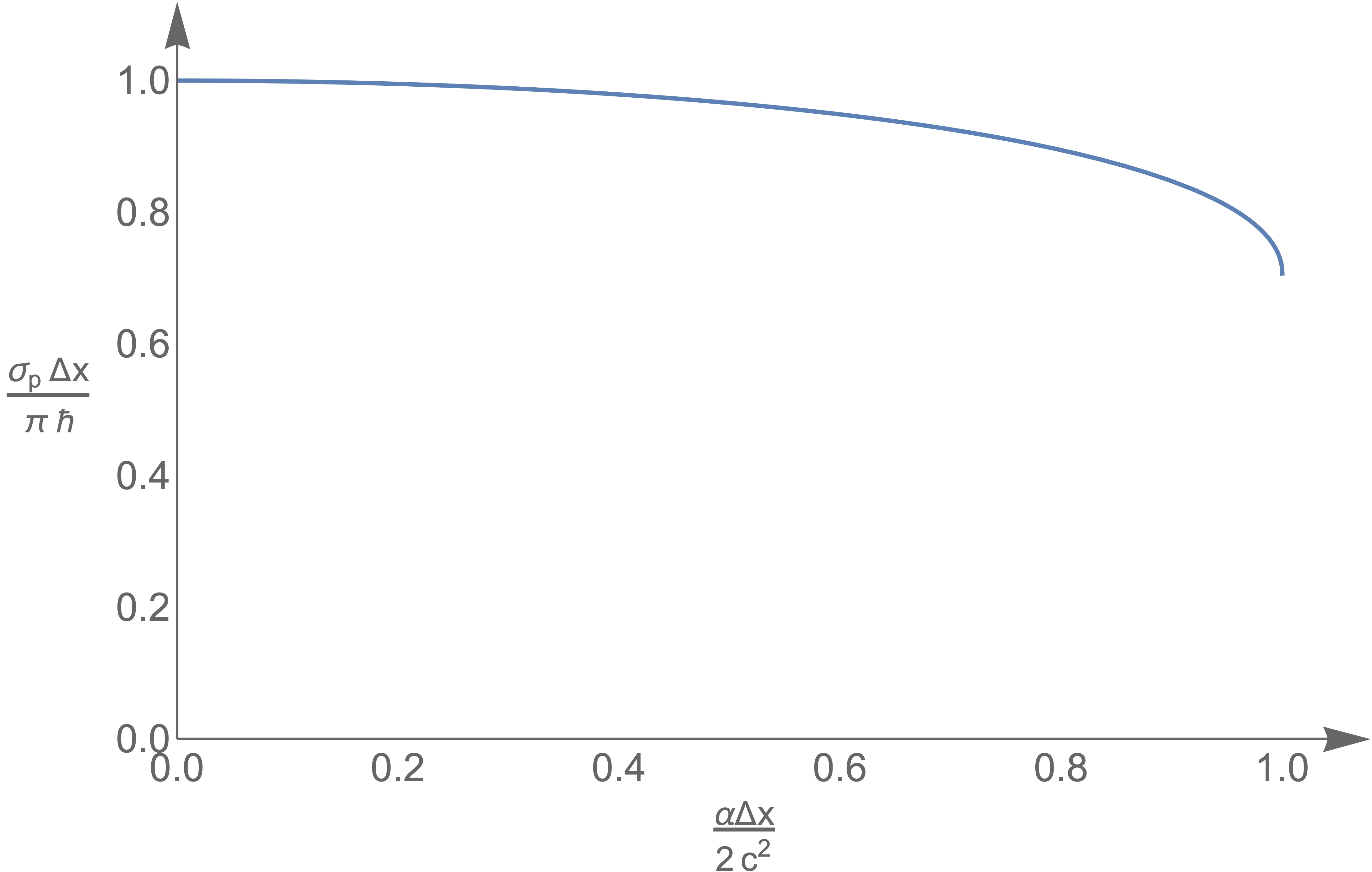}
\caption{The Extended Uncertainty Principle (\ref{EUPR}) for accelerated observers in terms of the rescaled position uncertainty  in units of $\pi\hbar$. In these units the uncertainty approaches a minimum value of $1/\sqrt{2}$. \label{fig:sol}}
\end{figure}

The covariant Laplacian along the direction of acceleration for the spatial part of the Rindler metric (\ref{Rmetric}) becomes
\begin{align}
\Delta_l&=\frac{\alpha}{c^2}\left(2 l \partial_l^2+ \partial_l\right).\label{laplacian} ,
\end{align}
so that the eigenvalue problem reads 
\begin{align}
l \psi_l''+\frac{\psi_l'}{2}+\tilde{\lambda}\psi_l&=0
\label{eigprob} ,
\end{align}
where $\psi_l$ stands for the part of the wave-function along the direction of acceleration, a prime denotes the derivative with respect to $l$, and $\tilde{\lambda}=2\lambda c^2/ \alpha$. The differential equation (\ref{eigprob}) has the general solution 
\begin{align}
\psi_l =c_1 \left[\cos\left(2\sqrt{\tilde{\lambda} l}\right)+c_2 \sin\left(2\sqrt{\tilde{\lambda} l}\right)\right] ,
\end{align}
with $c_1$, $c_2$ being 
constants. The constant $c_2$ and the eigenvalue $\tilde{\lambda}$ can be determined using the Dirichlet boundary conditions giving 
\begin{align}
&\cos\left(a\sqrt{\tilde{\lambda}}\right)+c_2\sin\left(a\sqrt{\tilde{\lambda}}\right)=0\label{c2} ,\\
&\cos\left[(a+\delta)\sqrt{\tilde{\lambda}}\right]+c_2\sin\left[(a+\delta)\sqrt{\tilde{\lambda}}\right]=0 ,
\label{eiganaprot}
\end{align}
where we have defined 
\begin{align}
a=\sqrt{2c^2/\alpha-\Delta x},~a+\delta=\sqrt{2c^2/\alpha+\Delta x}.
\end{align}
Solving \eqref{c2} for $c_2$ and plugging the result into \eqref{eiganaprot} one obtains that 
\begin{align}
\sin\left[(a+\delta)\sqrt{\tilde{\lambda}} - a\sqrt{\tilde{\lambda}} \right] = \sin\left(\delta \sqrt{\tilde{\lambda}}\right) = 0 ,
\end{align}
which is fulfilled, if the eigenvalues are given by 
\begin{align}
\label{eigR}
\lambda_n&=n^2\pi^2\frac{\alpha}{2c^2\delta^2}.
\end{align}
Using (\ref{Dpgen}) and (\ref{eigR}), the EUP for Rindler spacetime reads 
\begin{align}
\label{EUPR}
\sigma_p\Delta x \geq\pi\hbar\frac{\frac{\alpha\Delta x}{2c^2}}{\sqrt{1+\frac{\alpha\Delta x}{2c^2}}-\sqrt{1-\frac{\alpha \Delta x}{2c^2}}} ,
\end{align} 
which is an exact formula plotted in Fig. \ref{fig:sol}. While looking at the formula \eqref{EUPR} it is worth noticing that the uncertainty never reaches zero 
although it is monotonically decreasing with increasing $\Delta x$ and it features a minimum value of $1/\sqrt{2}$ in units of $\hbar/2$ where $\Delta x=l_0.$ Finally, for the sake of comparison with the common form of the EUP (presented  for example in Ref. \cite{GUPReview}), one can Taylor expand (\ref{EUPR}) for small values of $\alpha \Delta x/(2c^2)$ to get
\begin{align} 
\label{EUPRexp}
\sigma_p\Delta x \gtrsim\pi\hbar\left(1-\frac{\alpha^2(\Delta x)^2}{32c^4} + O\left[\left(\frac{\alpha^2(\Delta x)^2}{2c^4}\right)^2\right] \right) .
\end{align}

\section{EUP for Friedmann spacetime} 
\label{EFried}

In this section we consider Friedmann universe with hypersurfaces of constant Schwarzschild-like time (in deSitter/ anti-deSitter space this slicing corresponds to static coordinates). The corresponding spatial metric becomes
\begin{align}
\D s^2=\frac{\D r^2}{A(r,t_0)}+r^2\D \Omega^2\label{Fmet} ,
\end{align}
where
\begin{align}
A(r,t_0)=1-\frac{r^2}{r_H^2(r,t_0)}
\end{align}
with the apparent horizon 
\be
r_H^2=\frac{c^2}{H^{2}+\frac{K c^2}{a^2}} , 
\ee
the Hubble-parameter $H,$ the scale factor $a,$ the curvature index $K$, and the metric of the two sphere $\D \Omega.$ Note that in this approach the homogeneity of the universe is broken, putting an observer at the center of symmetry. Now, the universe is isotropic, though with respect to just one point, and not with respect to every point as it happens in maximally symmetric spaces.

\begin{figure}[!htb]
\centering
\includegraphics[width=.83\linewidth]{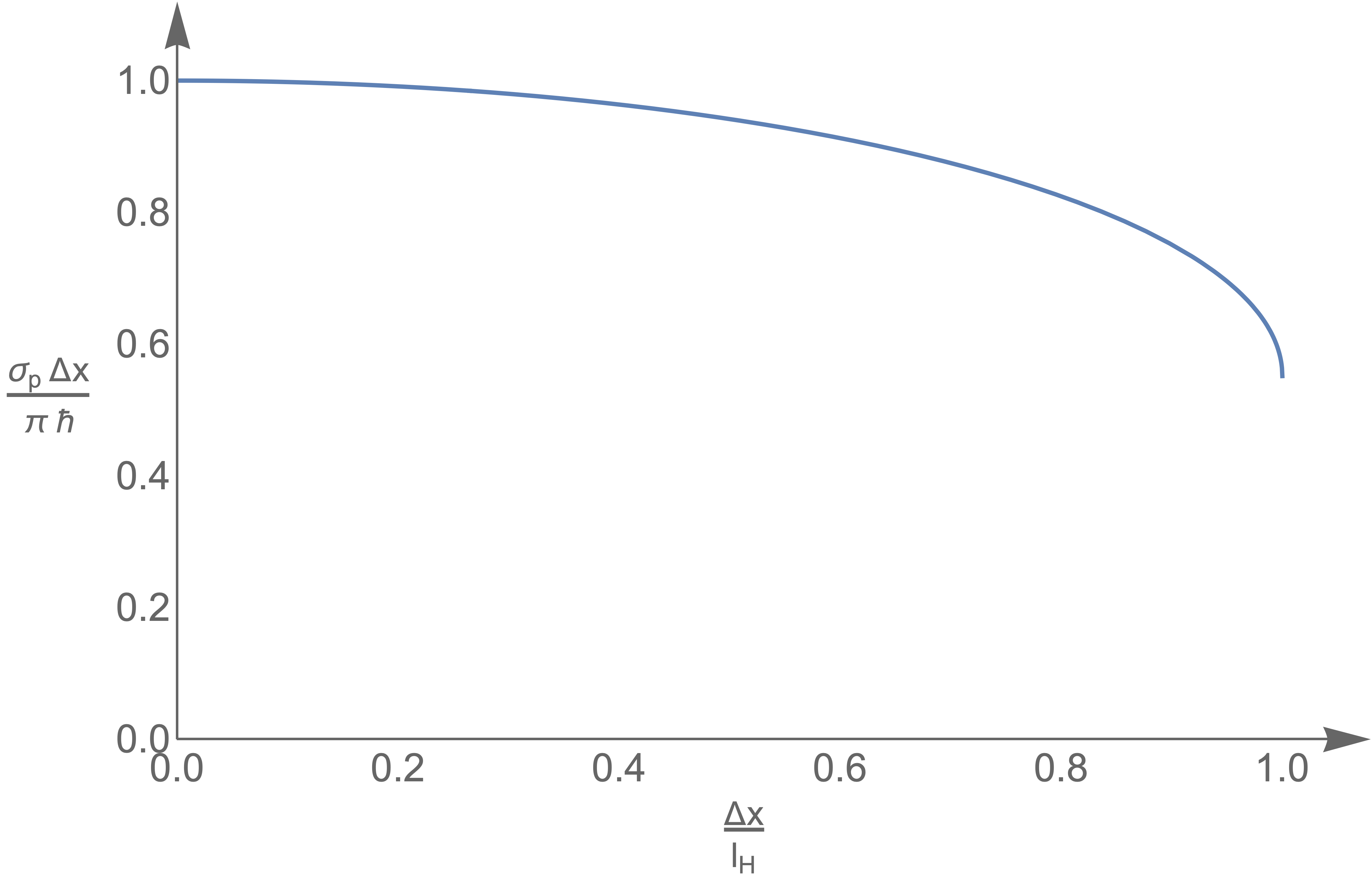}
\caption{The Extended Uncertainty Relation (\ref{EUPF}) for Friedmann background with manifest horizon in terms of the rescaled position uncertainty in units of $\pi\hbar$. In these units the uncertainty approaches a minimum value of $\sqrt{3}/\pi.$
\label{fig:sol2}}
\end{figure}

Matching the spherical symmetry of the metric, the domain to which the wave function is restricted is a geodesic ball $B_{\Delta x}$ of radius $\Delta x$ around the origin.
The covariant Laplacian along the radial direction for any spherically symmetric metric of the form \eqref{Fmet} reads 
\begin{align}
\Delta_r&=A(r) \partial_r^2+\left[\frac{2A(r)}{r}+\frac{A'(r)}{2}\right] \partial_r . 
\label{laplacian2}
\end{align}
Correspondingly, the radial eigenvalue problem becomes
\begin{align}
(1-\hat{r}^2) \psi_{\hat{r}}''+\left(\frac{2}{\hat{r}}-3\hat{r}\right)\psi_{\hat{r}}'+\tilde{\lambda}\psi_{\hat{r}}=0 ,
\label{eigprob2}
\end{align}
where $\psi_{\hat{r}}$ stands for the part of the wave-function along the rescaled radial coordinate $\hat{r}=r/r_H$ and $\tilde{\lambda}=r_H^{2}\lambda$. The differential equation (\ref{eigprob2}) has the general solution
\begin{align}
\psi_{\hat{r}}=\frac{\sqrt[4]{1-\hat{r}}}{\hat{r}}\left[c_1 P_a^{\frac{1}{2}}(\hat{r})+c_2 Q_a^{\frac{1}{2}}(\hat{r})\right] ,
\end{align}
where $P$ and $Q$ are associated Legendre polynomials of the \nth{1} and the \nth{2} kind, and we 
defined $a=\sqrt{1+\tilde{\lambda}}-1/2$. Taking into account the Dirichlet boundary conditions, we obtain the relations 
\begin{align}
c_1 P_a^{\frac{1}{2}}(0)+c_2 Q_a^{\frac{1}{2}}(0)&=0\label{c22} ,\\
c_1 P_a^{\frac{1}{2}}(\Delta \hat{x})+c_2 Q_a^{\frac{1}{2}}(\Delta\hat{x})&=0\label{eiganaprot2}, 
\end{align}
where $\Delta\hat{x}=\Delta x/r_H.$ Solving \eqref{c22} for $c_2$ and plugging into \eqref{eiganaprot2}, we obtain the condition 
\begin{align}
P_a^{\frac{1}{2}}(0)Q_a^{\frac{1}{2}}(\Delta\hat{x}) -P_a^{\frac{1}{2}}(\Delta \hat{x}) Q_a^{\frac{1}{2}}(0)&=0,
\end{align}
which after using the definition of $\tilde{\lambda}$, yields the EUP for the Friedmann spacetime
\begin{align}
\label{EUPF}
\sigma_p \Delta x \geq \hbar\frac{\Delta x}{r_H}\sqrt{\left(\frac{\pi}{2\arctan{f(\Delta x)}-\pi/2}\right)^2-1} ,
\end{align}
where
\begin{align}
\label{FdxF}
f(\Delta x)=\sqrt{\frac{1-\Delta x/r_H}{1+ \Delta x/r_H}} .
\end{align} 
The plot of (\ref{EUPF}) can be seen in figure \ref{fig:sol2}. Expanding (\ref{EUPF}) for small values of $\Delta x/r_H$ gives the standard form of such an EUP 
\begin{align}
\sigma_p\Delta x \gtrsim \pi\hbar\left(1-\frac{3+\pi^2}{6\pi^2}\frac{(\Delta x)^2}{r_H^2} + O\left[\left(\Delta x/r_H\right)^4\right] \right) 
\label{EUPFexp} .
\end{align}

\section{Relation to Hawking radiation \label{HawkingRadiation}}

The global minimum of momentum uncertainty entails an interesting implication. As we can assign the temperature
\begin{align}
T_{\sigma_p}=\frac{\sigma_p c}{k_B}
\end{align}
to the uncertainty of momentum, a global minimum of momentum uncertainty can be interpreted as global temperature or temperature of space-time. For the given horizons this temperature has the form
\begin{align}
T_{\sigma_p,min}=T_H \lim_{\Delta \tilde{x}\rightarrow 1} g(\Delta \tilde{x})
\end{align}
with the Hawking temperature of the respective horizons $T_H,$ $\Delta \tilde{x}=\Delta x/l_0$ for Rindler and $\Delta \tilde{x}=\Delta\tilde{x}/r_H$ for Friedmann space-time, respectively, and a function $g(\Delta \tilde{x})$ which basically possesses a limit of the order of 
$\sqrt{2}\pi^2$ for Rindler and $2\pi/\sqrt{3}$ for Friedmann universes respectively for horizon sized uncertainties ($\Delta \tilde{x}=1$).

Thus, the existence of a minimum of the momentum uncertainty can be understood as a different manifestation of 
Hawking radiation.

\section{Influence on black hole thermodynamics \label{BHthermo}}

The temperature of a Schwarzschild black hole can be derived heuristically using the standard HUP
\begin{align}
\sigma_p(\Delta x)\sim \frac{\pi \hbar}{\Delta x}.
\end{align}
Setting the uncertainty in position equal to the Schwarzschild radius $r_s$ with the black hole mass $M$ (in fact, $\Delta x$ is basically the radius, so the real uncertainty is twice this value), one can relate the Hawking temperature to the standard deviation of momentum as
\begin{align}
T_H^{(0)}=\frac{c}{k_B}\frac{\sigma_p(r_s)}{4\pi^2}=\frac{c \hbar}{k_B}\frac{1}{4\pi r_s}.
\end{align}
Using the first law of thermodynamics
\begin{align}
\frac{\D S}{\D E}=T^{-1}
\end{align}
with 
\be 
E=2 r_s \frac{c^4}{4G} = r_s F_{max} ,
\ee
where $F_{max}$ is the maximum force \cite{Gibbons,Schiller,PLB15},  
the entropy can be integrated to give
\begin{align}
S^{(0)}_{BH}=\frac{k_B c^3}{\hbar G}\pi r_s^2=\frac{k_B A_s}{4 l_p^2}
\end{align}
where $A_s$ and $l_p$ are the area of the Schwarzschild horizon and the Planck length respectively.

Taking into account the uncertainty relations obtained in Sections \ref{ERind} and \ref{EFried},  the departure from the standard Bekenstein entropy and Hawking temperature can be calculated. This will be done for the asymptotic form of the EUP with horizons and the exact relations \eqref{EUPR} and \eqref{EUPF} obtained above.

\subsection{Asymptotic form\label{asForm}}

As we can conclude from the previous results (\ref{EUPRexp}) and (\ref{EUPFexp}), the asymptotic form of the EUP for a background space-time which contains a horizon of radius $r_{hor}$ reads 
\begin{align}
\sigma_p\sim \frac{\pi\hbar}{\Delta x}\left(1+\beta_0 \frac{\Delta x^2}{r_{hor}^2} + O[(r_s/r_{hor})^4] \right) ,
\end{align} 
and leads to the Hawking temperature
\begin{align}
T_{H,as}=T_H^{(0)}\left(1+\beta_0 \frac{r_s^2}{r_{hor}^2} + O[(r_s/r_{hor})^4]  \right)
\end{align}
which yields an entropy
\begin{align}
S_{BH,as}&=\frac{ \pi k_B r_{hor}^2}{\beta_0 l_p^2}\log\left(1+\beta_0\frac{r_s^2}{r_{hor}^2} + O[(r_s/r_{hor})^4] \right)\\
           &\simeq S_{BH}^{(0)}\left(1-\frac{\beta_0}{2}\frac{r_s^2}{r_{hor}^2}+ O[(r_s/r_{hor})^4]\right)\\
		  &\simeq S_{BH}^{(0)}\left(1-\frac{\beta_0}{2}\frac{S_{BH}^{(0)}}{S_{hor}}+ O\left[\left(\frac{S_{BH}^{(0)}}{S_{hor}}\right)^2\right]\right) ,
\end{align}
where the horizon entropy of the background spacetime is equal to
\begin{align}
S_{hor}=\frac{\pi k_B r_{hor}^2}{l_p^2}.
\end{align}
Recall that $\beta_0 < 0$ for given spacetimes, so the total entropy of the black hole is increased - this is an effect of taking  canonical corrections due to some thermal fluctuations \cite{alphaposit2,alpha1,alphaposit}.

\subsection{Rindler space}

Applying the exact relation \eqref{EUPR}, the Hawking temperature of an accelerated black hole reads
\begin{align}
T_{H,R}=\frac{\hbar\alpha}{8 \pi c k_B}\left(\sqrt{1+\frac{\alpha r_s}{2c^2}}-\sqrt{1-\frac{\alpha r_s}{2c^2}} \right)^{-1}
\end{align}
which leads to the entropy
\begin{align}
S_{BH,R}=\frac{16\pi k_B}{3 l_p^2}\frac{c^4}{\alpha^2}\left[\left(1+\frac{\alpha r_s}{2c^2}\right)^{3/2}+\left(1-\frac{\alpha r_s}{2c^2}\right)^{3/2}\right]+S_0
\end{align}
with the integration constant $S_0$
\begin{align}
S_0=-\frac{32\pi k_B}{3 l_p^2}\frac{c^4}{\alpha^2}
\end{align}
chosen for proper normalization ($S_{BH,R}(r_s=0)=0$). Thus, the entropy becomes
\begin{align}
S_{BH,R}=\frac{16\pi k_B}{3 l_p^2}\frac{c^4}{\alpha^2}\left[\left(1+\frac{\alpha r_s}{2c^2}\right)^{3/2}+\left(1-\frac{\alpha r_s}{2c^2}\right)^{3/2}-2\right].
\end{align}
This entropy change encodes the entire non-perturbative influence of the Rindler horizon on an accelerated black hole. For small black holes ($\alpha r_s/2c^2\ll 1$) this result can be expanded to yield
\begin{align}
S_{BH,R}\simeq S_{BH}^{(0)}\left(1+\frac{S_{BH}^{(0)}}{16 S_R} + O\left[\left(S_{BH}^0/S_R\right)^2 \right] \right)
\end{align}
with the entropy of the Rindler horizon $S_R$ which is, of course, the result for the calculation in the asymptotic form.

\begin{figure}[!htb]
\centering
\includegraphics[width=.83\linewidth]{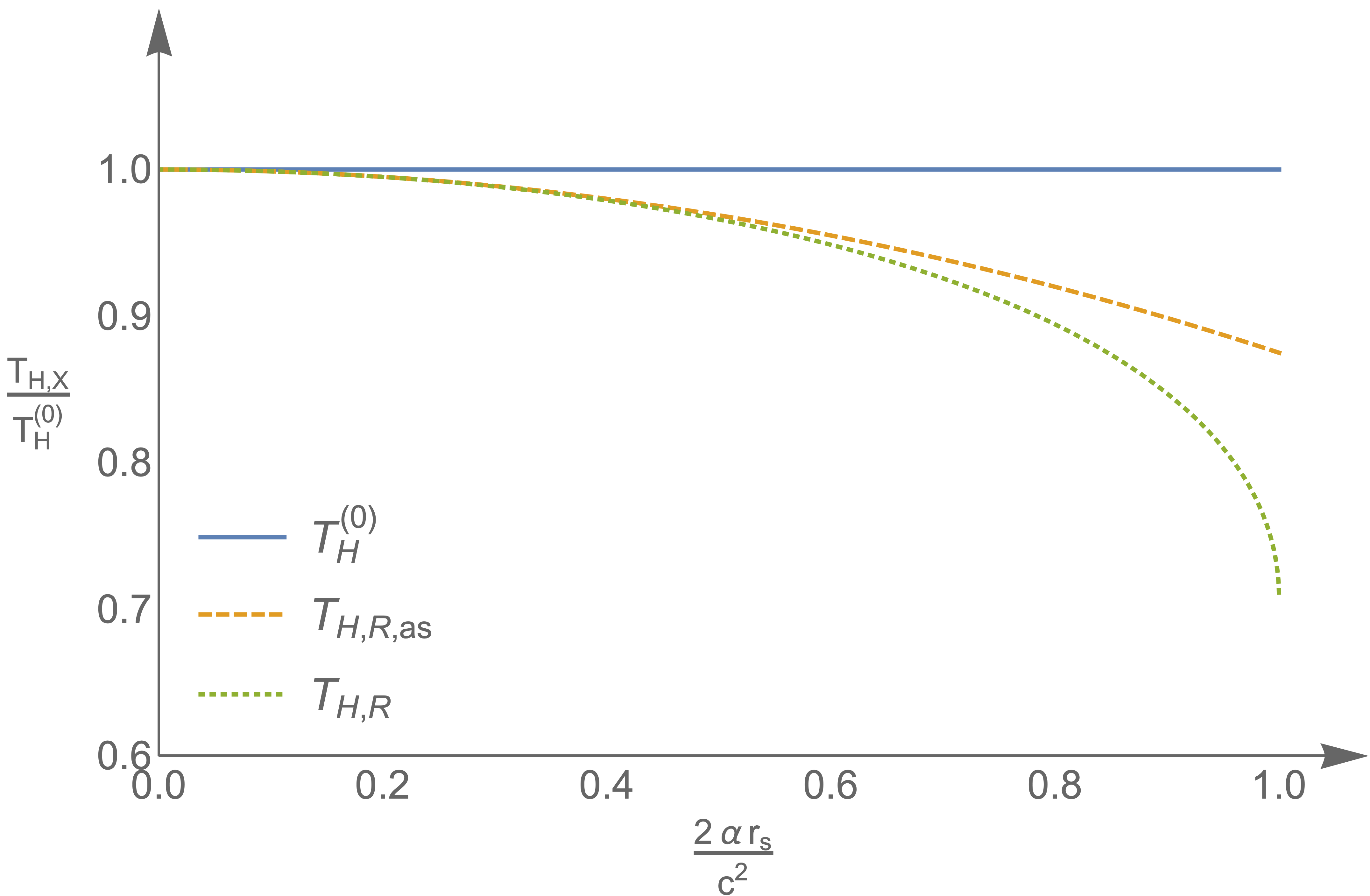}
\caption{The temperature of an accelerated black hole in units of the standard Hawking temperature as a function of the Schwarzschild horizon in units of the Rindler horizon distance $\alpha r_s/2c^2$ for fixed acceleration $\alpha$ in comparison to the asymptotic result.
\label{fig:TR}}
\end{figure}

\begin{figure}[!htb]
\centering
\includegraphics[width=.83\linewidth]{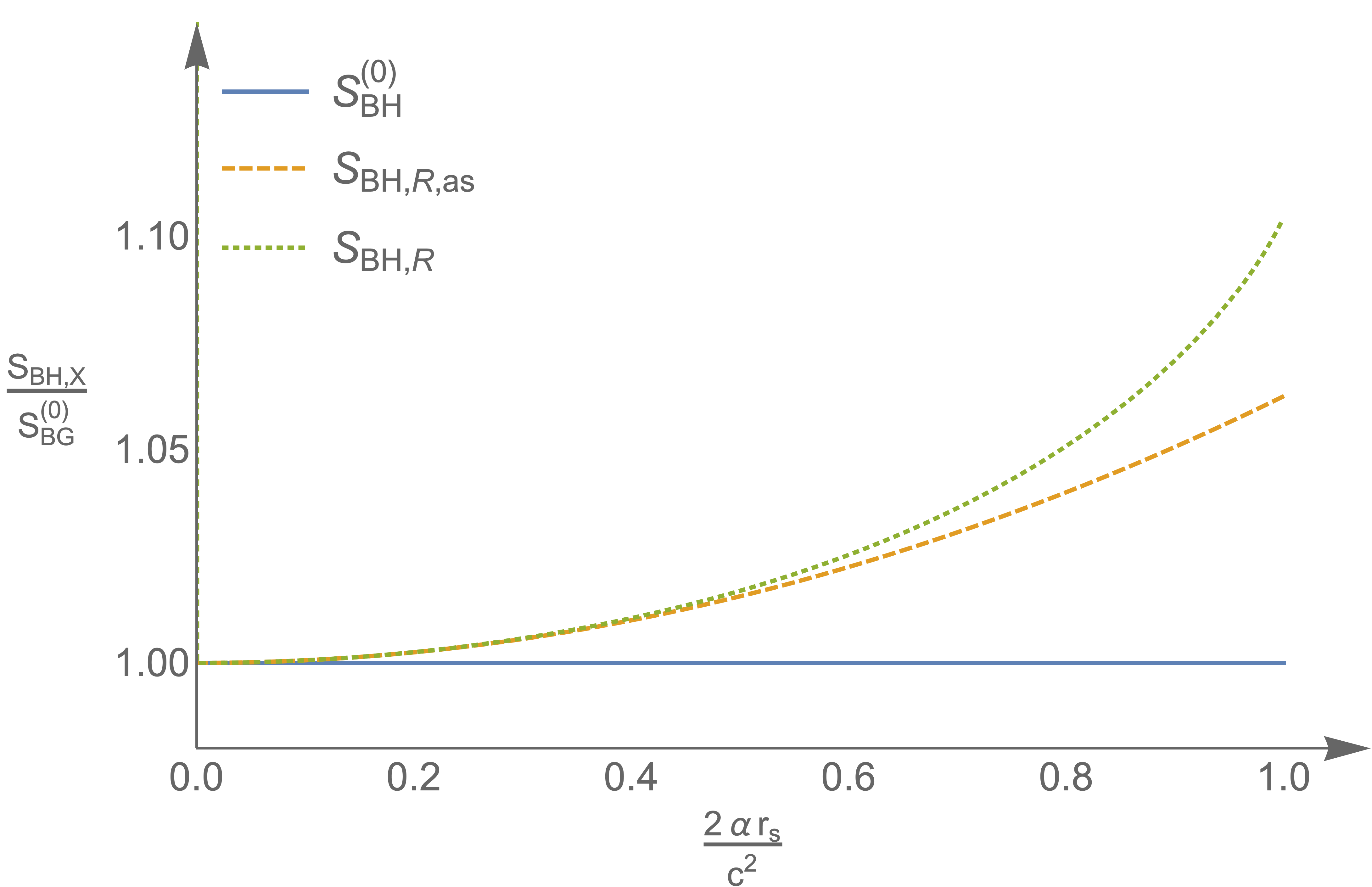}
\caption{The entropy of an accelerated black hole in units of the standard Bekenstein entropy as a function of the Schwarzschild horizon in units of the distance to the Rindler horizon $\alpha r_s/2c^2$ for fixed acceleration $\alpha$ in comparison to its asymptotic form.
\label{fig:SR}}
\end{figure}

Plots of the corresponding altered Hawking temperature and Bekenstein entropy are shown in Figs. \ref{fig:TR} and \ref{fig:SR}, respectively. As stated above, the presence of a Rindler horizon decreases the temperature of a black hole thus increasing its entropy. This effect is maximal when one uses the exact formulas.

\subsection{Friedmann space-time}

Analogously, the entropy of a black hole surrounded by a Friedmann horizon can be obtained. Correspondingly, the Hawking temperature becomes
\begin{align}
T_{H,F}=\frac{c\hbar}{k_B}\frac{1}{4\pi^2 r_H}\left(\sqrt{\left(\frac{\pi}{2\arctan f(r_s)-\pi /2}\right)^2-1}\right).
\end{align}
Unfortunately, the integration of the entropy cannot be done analytically. Therefore it will be given in its integral form
\begin{align}
S_{BH,F}=\frac{2\pi^2 k_B r_H}{l_p^2}\int \frac{\D r_s}{\sqrt{\left(\frac{\pi}{2 \arctan f(r_s)-\pi/2}\right)^2-1}}+S_0
\end{align}
with the integration constant $S_0,$ again, chosen in a way that $S_{BH,F}(r_s=0)=0.$

The expansion for small $r_s/r_H$ reads
\begin{align}
S_{BH,F}\simeq S_{BH}^{(0)}\left(1+\frac{3+\pi^2}{12\pi^2}\frac{S_{BH}^{(0)}}{S_H} + O\left[ \left(S_{BH}^0/S_H \right)^2 \right]\right) ,
\end{align}
where the Hubble-horizon entropy $S_H$ equals to the asymptotic result.

\begin{figure}[!htb]
\centering
\includegraphics[width=.83\linewidth]{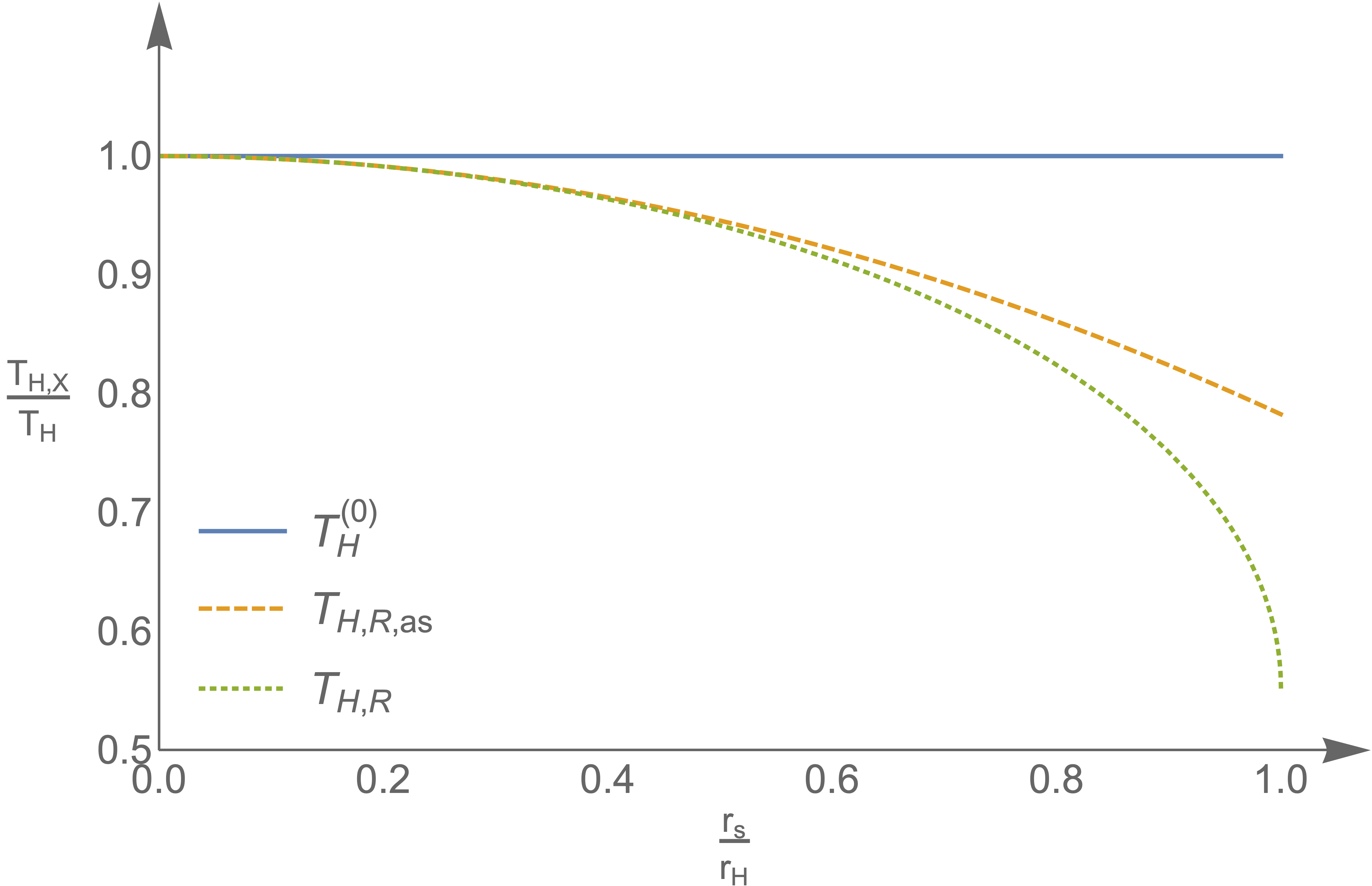}
\caption{The Hawking temperature of a black hole surrounded by a cosmological horizon in units of the standard Hawking temperature as a function of the Schwarzschild horizon in units of the cosmological horizon distance $r_s/r_H$ for a fixed horizon distance $r_H$ in comparison to the asymptotic result.
\label{fig:TH}}
\end{figure}

\begin{figure}[!htb]
\centering
\includegraphics[width=.83\linewidth]{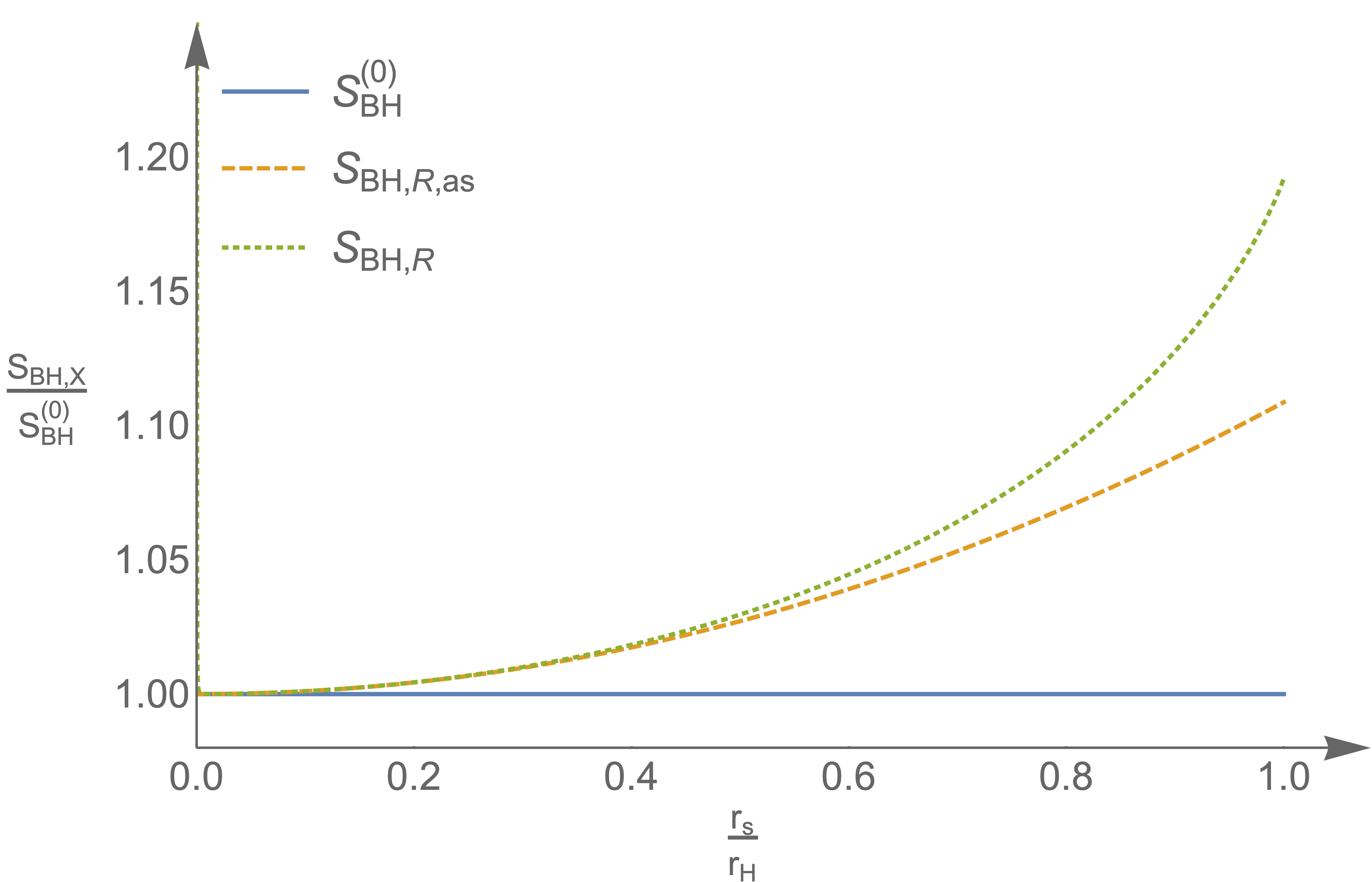}
\caption{The entropy of a black hole surrounded by a cosmological horizon in units of the standard Bekenstein entropy as a function of the Schwarzschild horizon in units of the distance to the cosmological horizon $r_s/r_H$ for fixed $r_H$ in comparison to its asymptotic form.
\label{fig:SH}}
\end{figure}

Plots of the modified Hawking temperature and Bekenstein entropy are shown in Figs. \ref{fig:TH} and \ref{fig:SH} respectively, where the latter was computed numerically. Analogously to the Rindler case, the presence of the horizon decreases the temperature and increases the entropy. As for the Rindler horizon, the application of the exact relation results in a considerable amplification of this effect.

\section{Summary}
 \label{Sum}

The influence of the Rindler acceleration $\alpha$ and the cosmological horizon $r_H$ 
on the uncertainty relation has been derived. The solutions recognize the horizons at $l_0=2c^2/\alpha$ and $r_H$ as maximal position uncertainties and recover the usual Heisenberg uncertainty principle for $\alpha \to 0$ and $r_H \to \infty$, respectively. While crossing the horizons, the uncertainties become imaginary meaning that the position and momentum operators cease to be observables. 

In general, both cases show a very similar behaviour. This indicates that maximum lengths by external horizons leave a very particular imprint on the uncertainty relation: not only does the uncertainty become imaginary, but it also does not go to zero even though it continuously decreases thus providing a natural momentum cut-off. This cut-off has values $\sigma_p\geq \hbar\pi\alpha/\sqrt{8}$ and $\sigma_p\geq \sqrt{3}\hbar/r_H$
for Rindler and Friedmann spaces respectively. In contrast, a maximum length by topology as derived in Ref. \cite{Schuermann2018} is given by \eqref{Kgeom}. For positive $K$ it becomes zero before turning imaginary, so the momentum does not get restricted which is sensible because the wavelength can cover a closed universe several times without any problem.

This result implies that the existence of horizons constrains the momentum uncertainty which can be interpreted as assigning a temperature to a spacetime that contains a horizon just as it is done in terms of Hawking radiation. Consequently, the minimum momentum uncertainty is of the order of the Hawking temperature. We can then identify the presence of Hawking radiation with this particular influence on the uncertainty relation thus understanding the latter as a manifestation of the former.

Finally, the effects of Rindler and cosmological horizons on black hole thermodynamics have been analysed heuristically for the asymptotic and exact momentum uncertainties derived before thereby showing that the temperature is decreasing while the entropy is increasing. In particular, the application of exact solutions yields a considerable amplification of this effect. 


\section*{Acknowledgements}

MPD appreciates the discussions with Ana Alonso-Serrano and Hussain Gohar. FW thanks for useful hints by Samuel Barroso-Bellido.

\end{document}